\documentclass[twocolumn,pre,floatfix,superscriptaddress]{revtex4}
\usepackage{graphicx}
\usepackage{amsmath}
\usepackage{latexsym}
\usepackage{epstopdf}
\usepackage{amsmath}
\usepackage{amssymb,amsmath}
\usepackage{color}

\newcommand{\beq}{\begin{equation}}
\newcommand{\eeq}{\end{equation}}

\begin{document}
\title{First passage time processes and subordinated SLE}
\author{M. Ghasemi Nezhadhaghighi}
\affiliation{Department of Physics, Sharif University of Technology, Tehran, P.O.Box: 11365-9161, Iran}
\author{M.~A.~Rajabpour}
\affiliation{SISSA and INFN, \textit{Sezione di Trieste},  via Bonomea 265, 34136 Trieste, Italy}
\author{S. Rouhani}
\affiliation{Department of Physics, Sharif University of Technology, Tehran, P.O.Box: 11365-9161, Iran}

%



\begin{abstract}

We study the first passage time processes of the anomalous diffusion on the self similar curves in two dimensions. The scaling properties of the mean square displacement and mean first passage time of the fractional Brownian motion and subordinated walk on the different  fractal curves (loop erased random walk, harmonic explorer and percolation front) are derived. We also define
natural parametrized subordinated Schramm Loewner evolution (NS-SLE) as a mathematical tool
that can model diffusion on fractal curves. The scaling properties of the mean square displacement
and mean first passage time for NS-SLE are obtained by numerical means. 

\end{abstract}
\maketitle
\section{Introduction} 

The anomalous or non-Fickian diffusive transports have attracted a lot of interest in the past few years. There is a wide range of heterogeneous or pre-asymptotic systems  in the fields of physics, astronomy, biology, chemistry, and economics, where anomalous diffusion occur \cite{metzler}.  This phenomenon is observed in diffusion on fractal structures in geophysical and geological media \cite{fractdiff,Berkowitz}, charge transport in disordered and
amorphous semiconductors \cite{scher}, acceleration of particles inside a turbulent medium \cite{torb}, transport process in the biological systems \cite{bio}, and many other examples, for an extensive collection of references see \cite{metzler}. The well-known examples of anomalous diffusions are the continuous time random walk (sub-diffusive processes) \cite{Barkai} and L\'evy flight \cite{Levyflight}.
In this type of processes, the mean square displacement (MSD) obeys a power law equation with respect to the time with exponent $0< \nu < 2 $. Anomalous transport, especially the continuous time random walk and L\'evy flight can be studied within the fractional Fokker Planck equation approach \cite{FFPE}.

One of the most important characteristics in normal and anomalous diffusions is the first passage time (FPT), which is defined as the time needed the dynamic variable to cross a given threshold value for the first time \cite{stochastic,FPTD}. The FPT has been used to characterize diffusive processes in various systems such as the spreading of disease \cite{disease}, the passage
of polymers and DNA in sub-diffusive media and membranes \cite{DNAtransport}, the firing of neurons \cite{neuron}, animals searching for food \cite{animal} and L\'evy stable random motion \cite{Levyflight}. 

One of the interesting subjects of the diffusion problem is the study of FPT processes in fractal geometries such as percolating fronts, crack patterns, polymer chains, lightening paths, etc \cite{fractal,fracture,Voss,Sapoval,lightning}. In this work we are interested in simple fractal objects with fractal dimension $1<d_f<2$ without any branch point. Since the coordination number of all the points on the fractal is two it is easy to conclude that there should be lots of similarities between FPT processes in these systems and one dimensional systems. This was already discussed in \cite{Zoia} and the important rule of the length of the fractal objects in FPT processes were understood. In \cite{Zoia} the connection with the Schramm Loewner evolution (SLE) \cite{Schramm} were also discussed, especially the important rule of the definition of the length in the SLE studies \cite{Zoia,kennedy} were emphasized. In this work we will generalize the work done in \cite{Zoia}, in many different directions. To have an idea of fractal objects with fractal dimension ${1<d}_f<2$ we will study the scaling exponents of MSD and mean FPT (MFPT) of the diffusing particles on loop-erased random walk (LERW), harmonic explorer (HE) and percolation fronts (PF) on the upper half plane. We will study different random walkers such as fractional Brownian motion and subordinated walk on the fractal curves. The scaling properties of the MSD of the walker and MFPT will be discussed. Finally we give a novel method to study diffusion on fractal curves by using SLE. We show that all the scaling behaviors discussed for the discrete fractal curves can be rederived by using subordinated natural SLE which is the time changed Schramm Loewner evolution.\newline
The central aim of this work is to use the subordinated natural SLE to find a new connection between diffusion on the self similar traces and SLE as a growth process. Our analysis of MSD and MFPT are the key points of this connection.\newline
The paper is organized as follows: In the next section we will fix the notation and introduce the scaling relations for the diffusion problem on semi-1D fractal paths, where we measure the scaling exponents of MSD and MFPT, for  two sided diffusion and diffusion with waiting times on the self similar curves with fractal dimension $d_f$. 
In the third section we use the Schramm Loewner evolution (SLE) as a mathematical model to introduce new classes of diffusion processes (subordinated SLE). The  results of this section are compatible with the two-sided diffusion and diffusion with the waiting time on the discrete fractal paths.  
In the last section we conclude our findings. To be self explanatory we add three appendices explaining the details of our simulation methods. 

\section{First passage time in lattice fractal interfaces}

We begin by considering the diffusion problem on semi-1D random curves. For our purposes we restrict ourselves to the fractal curves with \textit{Hausdorff} dimensions $1<d_f<2$ \cite{fractal} that start from the origin and remain in the upper half-plane. To understand statistical properties of diffusion along such inhomogeneous paths it is important to first introduce diffusion problem in 1D case.

\subsection{First passage time statistics in one dimensional domain}

Consider a one-dimensional diffusion $X_t$ with dynamics
\begin{equation} \label{1Ddynamic}
dX_t =a(X_t)dt +\sigma  dW_t,  
\end{equation} 
where $W_t$ is a one dimensional stochastic process. The interval for the solution of Eq. (\ref{1Ddynamic}) is defined as closed on the left hand side $x_a=0$ and open on the right hand side $x_b=R$. These special choices force diffusing particle to move only in $x\geq0$. 

An interesting problem in the theory of stochastic processes is finding the time that a particle reaches a certain level. The problem of finding this time is called first passage time (FPT) \cite{stochastic,FPTD}. The first passage time is the time ${\tau }_r$ taken for the process having started from $x=0$ to be reached in $x=r$ \cite{FPTD}
\begin{equation} \label{FPT} 
{\tau }_r=\inf \lbrace{t>0|X_t=r}\rbrace,
\end{equation}
where the infimum for every subset $S$ of real numbers is denoted by $\inf \{S\}$ and is defined to be the biggest real number that is smaller than or equal to every number in $S$.

Clearly FPT is a random variable which varies from one sample of $X_t$ to another one. In general we are interested in those processes where the two statistical variables $\langle{X^2_t}\rangle$ and $\langle{\tau _r} \rangle $ have scaling behavior
\begin{eqnarray} \label{GenscalingX2} 
\langle{X^2_t}\rangle \propto \ t^{\nu },\hspace{1cm}
{\langle{\tau }_r}\rangle \propto r^{\beta } .
\end{eqnarray}

One example of Eq. (\ref{1Ddynamic}) is $a=0$, $\sigma =1$ and $W_t={|B}^H_t|$, where $B^H_t$ is the fractional Brownian motion (fBm)  process. Note that, fBm with Hurst index $0<H<1$ is the only self-similar Gaussian process with stationary of increments \cite{FBM}. The correlation function of fBm is
\begin{equation} \label{fBmcorrelation} 
\langle{B^H(t)B^H(s)}\rangle \sim [{|t|}^{2H}+{|s|}^{2H}-{|t-s|}^{2H}].
\end{equation} 
Absolute value of  $B^H_t$ is used to force diffusing particle to move only in the region $x\geq0$. The particle's position $X_t={|B}^H_t|$  is a stochastic variable with 
\begin{equation} \label{fBmscalingX2} 
\langle{X^2_t}\rangle=\langle{|B^H_t|}^2\rangle\ \sim t^{2H}. 
\end{equation} 
The scaling relation in Eq. (\ref{fBmscalingX2}) is in agreement with Eq. (\ref{GenscalingX2}) and shows that $\nu =2H$ \cite{Ding}. Diffusion is said to be anomalous if $\nu \neq 1$, where $0<\nu <1$ ($1<\nu <2$) indicates the sub-diffusive  (super-diffusive)  behavior.

It is easy to determine the exponent $\beta =1/H$ analytically for the mentioned boundary condition. To find the scaling parameter $\beta$, consider a random process $X_t$ as $X_{t/{\lambda} ^{1/H}}=\frac{1}{\lambda} X_t$. Now one can replace $r\rightarrow \lambda r $ in Eq. (\ref{FPT}) to find $\tau_{\lambda r}$. It is straightforward to show that  $\tau_{\lambda r}$ is the same as ${\lambda^{1/H}}\tau_{r}$ in the distributional sense. By this scaling argument, one can observe that  MFPT is given by
\begin{equation} \label{fBmscalingFPT} 
{\langle{\tau }_r}\rangle \propto r^{1/H }. 
\end{equation}
It is worth mentioning that by choosing $W_t=|B_t|$, where $B_t$ is a Brownian motion (fBm with $H=\frac{1}{2}$), one can use the Fokker-Planck (FP) equation, which describes the space-time evolution of the probability density function (pdf) of $X_t$, to find FPT distribution. More details can be found in \cite{hinkel}.

Although studying diffusion in one dimensional systems is interesting for its own sake there are many examples of diffusion in self-similar interfaces with fractal dimensions $1<d_f<2$. We will generalize the above arguments to self-similar interfaces with arbitrary fractal dimension and will study the statistical properties of the diffusion on the fractal curves.

To motivate our method of measurement of the scaling parameters $\nu $,  $\beta $  for diffusing particle on the fractal curves we first consider models on a lattice domain, e.g., the loop-erased random walk, harmonic explorer and the percolation explorer process.

\begin{itemize}
\item[a. ]\textit{The loop-erased random walk}
\end{itemize}
The LERW on the square lattice domain is a random walk with erased loops when they appear. This process is stopped when it reaches a given length. To produce LERW curves started from the origin and conditioned to be in the upper half plane one can use reflecting boundary condition on the real axis for the random walker. The fractal dimension of LERW is  5/4 \cite{confinvarLERW}.

\begin{itemize}
\item[b. ]\textit{An explorer processes}
\end{itemize}
Explorer processes (EP) on honeycomb lattice such as percolation front with $d_f=\frac{7}{4}$ and harmonic explorer with $d_f=\frac{3}{2}$ are used as other classes of fractal interfaces \cite{bauer,kager}. To construct an EP path with a fixed number of steps $N$, we used from a class of explorer processes on the honeycomb lattice. This process is named \textit{overruled harmonic explorer } \cite{celani}. For numerical analysis, we simulated this process on the extremely large rectangular domain, where it can approximate the upper half plane (see appendix. A).

 Although other studies such as \cite{Zoia} have also used self similar traces to study the FPT problem they have not characterized the scaling relations in the measurable quantities such as MSD and MFPT for the general diffusion processes e.g. two sided diffusion and diffusion with the waiting time which we have studied in our simulations.

\subsection{Two sided Diffusion on the fractal paths}

In this subsection we study the statistical properties of the diffusing particles along self-similar curves. An interesting problem in this direction is the determination of the scaling exponents of the random displacement. 

For one dimensional domain with reflecting boundary condition, we mentioned in section 2.1 that the MSD and the MFPT according to Eq. (\ref{fBmscalingX2}) and Eq. (\ref{fBmscalingFPT}), obey scaling laws with the exponents $\nu=2H$ and $\beta=1/H$. In the following we explain how to use the discrete random walk model to simulate stochastic process $X_t$ on 1D domain with reflecting boundary condition on $x=0$ and we then favourably apply this model to random process on the discrete fractal curves.

First we consider the random walker on the one dimensional discrete lattice. This random walker started from the position $x=0$ at $t_0=0$ and at the time $t_n=n\delta t$ moves one step to right (left) when $\lbrace|B^H_{t+\delta t}|-|B^H_t|\rbrace>0$ ($\lbrace|B^H_{t+\delta t}|-|B^H_t|\rbrace<0$). The normal random walk (discrete version of Brownian motion) corresponds to $H=1/2$. This random process corresponds to two sided diffusion on 1D domain. Following the idea presented in one dimension, we obtain the statistics of two sided diffusion on the fractal curves. To this aim we consider a random walker with position coordinates $X_n$ and $Y_n$ for the $n^{th}$ walk where $(X_0,Y_0)=(0,0)$ is the start position and it moves back and forth along the discrete self-similar curve. In order to simulate random walker on the curve started from the origin and remains in the upper half plane we used from fBm process  $|B^H_t|$. Using this correlated stochastic process we define another stochastic process $S^H_n$ so that $S^H_{n+1}=S^H_n+1$ ($S^H_{n+1}=S^H_n-1$) when $\lbrace|B^H_{t+\delta t}|-|B^H_t|\rbrace>0$ ($\lbrace|B^H_{t+\delta t}|-|B^H_t|\rbrace<0$) with the initial value $S^H_0=0$. The random walk position can be defined by $X_n=x(S^H_n)$ and $Y_n=y(S^H_n)$ where $x(i)$ and $y(i)$ are the position components of the $i^{th}$ point of the curve. 

We now study numerically scaling dependence of $\langle{R^2_n}\rangle=\langle{X^2_n}+{Y^2_n}\rangle$ and $\langle{\tau_r}\rangle$ to $n$ and $r$ for many random walkers moving along such self similar one-dimensional objects. Especially we study the scaling forms of $\langle{R^2_n}\rangle$ and $\langle{\tau_r}\rangle$. 

To study the scaling laws in two sided diffusion, it should be mentioned here that the scaling exponents $\nu(H,d_f)$ and $\beta(H,d_f)$ in two sided diffusion on the fractal curves are in general a function of \textit{Hurst} parameter $H$ and geometrical dimension $d_f$. 
Within the FPT statistics of one dimensional diffusing particle approach, one can consider the self-similar curve as a one dimensional non-straight line with length $l$. For this semi-1D object using Eq. (\ref{fBmscalingFPT}) one can observe that $\langle \tau_l\rangle \propto l^{1/H}$. On the other hand for a fractal curve (with length $l$) inside a circle there is a scaling law $l\propto r^{d_f}$ where $r$ is the radius of the circle. Under these assumptions, the MFPT reads as $\langle \tau_r\rangle \propto r^{d_f/H}$.

The very same method can be applied to scaling law of MSD. Same as before we consider the fractal curve as a semi-1D object. There is a scaling relation $\langle l_n ^2\rangle \propto n^{2H}$ (similar to Eq. \ref{fBmscalingX2}) for the position of diffusing particle $l_n$ after $n$ walks along such semi-1D curve. In addition, the scaling relation $l_n\sim R_n^{d_f}$ for the fractal curve is well-known, where $R_n$ is the radius of the semi-circle enclosed $l_n$-th walks. We therefore obtain a universal law   $\langle{R^2_n}\rangle \propto n^{2H/d_f}$  for MSD of two sided diffusing particles on the fractal curves. Using these arguments we expect 

\begin{equation} \label{twosidedscalingX2} 
\nu(H,d_f)=2H/d_f ,\hspace{1cm} \beta(H,d_f)=d_f/H,
\end{equation}
These equations are in agreement with the scaling exponents in the Eqs. (\ref{fBmscalingX2}) and (\ref{fBmscalingFPT}) in the $d_f\rightarrow 1$ limit. 
The above results for $H=\frac{1}{2}$ recover the predictions in \cite{Zoia}. Table ~\ref{tab1} summarizes our numerical results for the two scaling exponents $\nu$ and $\beta$, where it shows that our results are well compatible with the predictions  in Eq. (\ref{twosidedscalingX2}) for two sided diffusion on the fractal curves. In our measurements we used from 50000 fractal curves and 10 independent realizations of random process $S^H_n$ per curve  for each numerical test.

\begin{table}[htp]
\begin{center}
\begin{tabular}{ccccc}\hline\hline
Model &   $\nu(H=0.8)$  & $\beta(H=0.8)$ &  $\nu(H=0.9)$  & $\beta(H=0.9)$\\
\hline
LERW &$1.280 \pm {0.001}$ & $1.56 \pm {0.01}$&$1.437 \pm {0.005}$ & $1.385 \pm {0.005}$\\
HE  &$1.066 \pm {0.001}$ & $1.88 \pm {0.01}$&$1.195 \pm {0.005}$ & $1.670 \pm {0.003}$\\
PF  &$0.92 \pm {0.01}$ & $2.17 \pm {0.01}$&$1.029 \pm {0.003}$ & $1.945 \pm {0.005}$    

\\ \hline\hline
\end{tabular}
\end{center}
\caption{\label{tab1} Numerical values of the scaling exponents $\nu$ and $\beta$ for the two sided diffusion on the fractal curves. To measure these exponents we used from correlated process $S^H_n$ with $H$=0.8 and 0.9.}
\end{table}

\subsection{Diffusion with waiting times on the fractal curves }

In this subsection we study anomalous motion of a free particle with waiting time on the self-similar curve. In the preceding section, we have obtained statistics of walkers moving back and forth randomly along self-similar discrete curves. The physical time between two consecutive steps of walks in the previous examples is equal to a constant $\Delta t$. 

We can also consider the random time elapsing between two consecutive jumps of a diffusing particle. The particle starts from the origin and trapped in site $n$  for some random time. These positive random waiting times $\tau_n$ are identically distributed random variables each having the same probability density function $\psi(\tau)$ \cite{Barkai,Berkowitz}. 
The role of the waiting time forces us to identify operational time $S^\alpha_{t_n}= \sum_n \tau_n$ and physical time $t_n=n\Delta t$. The physical time  $n$  is always accelerated against the strictly increasing random time $S^\alpha_{t_n}$. This is so, the random time $S^\alpha_{t_n}$ is called subordinator. As mentioned in \cite{Sokolov,Magdziarz,Janicki} it is described as

\begin{equation} \label{subordinator} 
S^\alpha_{t_n}={inf}\lbrace\tau_m :U(\tau_m)>t_n\rbrace,
\end{equation}
where  $U(\tau_m)$ is $\alpha$-stable subordinator ($0<\alpha<1$) and $\tau_m=m\Delta \tau$. The random process $S^\alpha_{t_n}$ is called the inverse-time $\alpha$-stable subordinator. 
The above process has neither stationary nor independent increments but it is easy to show
that we have distributional scaling $S^\alpha_{\lambda t}=\lambda^{\alpha}S^\alpha_{t}$ which leads us to the
following symmetry for the subordinated Brownian motion
\begin{equation} \label{subordinated_brownian_scaling} 
B(S^\alpha_{ct})=B(c^{\alpha}S^\alpha_{t})=c^{\alpha/2}B(S^\alpha_{t}).
\end{equation}
Although $B(S^\alpha_{t})$ is self similar with \textit{Hurst} exponent $\alpha/2$ it is not fractional Brownian motion
because it does not have a Gaussian distribution and it does not have stationary increments.
The process $S^\alpha_{t}$ is strictly increasing and it tends to infinity for $t\rightarrow\infty$ and so it is a good
process to consider as the time, whereas in the $\alpha\rightarrow1$ limit the subordinated time converges to the physical time. In our study we generated the subordinator $S^\alpha_{t_n}$ following the reference \cite{Magdziarz} (see appendix. B).

We study statistical properties of position coordinates $X_n=x(S^\alpha_{t_n})$ and $Y_n=y(S^\alpha_{t_n})$ for a subordinated walker with operational time $S^\alpha_{t_n}$ moves on the discrete self-similar curve with fractal dimension $d_f$ and position components $x(i)$ and $y(i)$. For this class of particle diffusion in fractal path, we again expect  universal scaling dependence of the MSD ($\langle{R^2_n}\rangle=\langle{X^2_n}+{Y^2_n}\rangle$) and the MFPT ($\langle{\tau_r}\rangle$) to the geometrical parameters $n$ and $r$  as

\begin{equation} \label{subordinatedscalingX2} 
\nu(\alpha,d_f)=2\alpha/d_f, \hspace{1cm}  \beta(\alpha,d_f)=d_f/\alpha.
\end{equation}
Our numerical results for $\alpha=0.8$ and $\alpha=0.9$ (see Table ~\ref{tab2})  are in good agreement with the scaling exponents in Eq. (\ref{subordinatedscalingX2}).

\begin{table}[htp]
\begin{center}
\begin{tabular}{ccccc}\hline\hline
Model &  $\nu(\alpha=0.8)$  & $\beta(\alpha=0.8)$ & $\nu(\alpha=0.9)$  & $\beta(\alpha=0.9)$ \\
\hline
LERW &$1.280 \pm {0.001}$ & $1.56 \pm {0.01}$&$1.440 \pm {0.001}$ & $1.39 \pm {0.01}$\\
HE  &$1.065 \pm {0.005}$ & $1.88 \pm {0.01}$&$1.200 \pm {0.001}$ & $1.67 \pm {0.01}$\\
PF  &$0.920 \pm {0.005}$ &$2.18\pm {0.01}$&$1.030 \pm {0.005}$ & $1.95 \pm {0.01}$   

\\ \hline\hline

\end{tabular}
\end{center}
\caption{\label{tab2} Numerical values of the scaling exponents $\nu$ and $\beta$ for subordinated diffusion on the fractal curves. To measure these exponents we used from subordination $S^{\alpha}_{t_n}$ with $\alpha$=0.8 and 0.9.}
\end{table}

\section{First passage time and Schramm Loewner evolution}

In the preceding section we studied scaling exponents of MSD and MFPT for some important examples of two sided and subordinated diffusion on the discrete self-similar curves in the upper half plane. In our study we used three statistical models, loop-erased random walk, harmonic explorer and percolation interfaces on the lattice.  

The scaling limit of the lattice models as the lattice spacing goes to zero corresponds to Schramm Loewner evolution. This mathematical model is defined in the complex plane and it was introduced by Schramm \cite{Schramm}. SLE is based on the Loewner equation  
\begin{equation} \label{Loewner_equation} 
\partial_t g_t(z)=\frac{2}{g_t(z)-\xi_t},
\end{equation}
where the real-valued function $\xi_t$ is called driving (or forcing) function, which determines all the
properties of SLE. Loewner showd that for any non-intersecting curve parametrized by a complex function $\gamma (t)$ in the upper-half plane $\mathbb{H}$, there exist a conformal map $g_t(z)$, which maps upper
half plane minus curve and the region which, is separated from infinity by the curve (\textit{hull}: $K_t$) $\mathbb{H} \setminus K_{t}$, to the upper half plane $\mathbb{H}$ \cite{Lawler0}.

Ordinary SLE  is the Loewner evolution with $\xi_t=\sqrt{\kappa}B_{t}$, where $B_t$ is the Brownian
motion with mean zero and $E[B_{t}B_{s}]=min(t,s)$ and also with diffusion constant $\kappa>0$ \cite{Schramm}. These properties ensure that the curve is conformally invariant. 

SLE$_\kappa$ is a random conformally invariant curve with the fractal dimension $d_f=1+\kappa/8$ ($0<\kappa<8)$ \cite{bauer,Cardy}. The scaling limits of LERW, HE and PF are SLE$_\kappa$ with $\kappa$ = 2, 4 and 6, respectively \cite{confinvarLERW,chramma,Smirnov}.

The SLE$_\kappa$ curve ($\gamma(t)$) is parametrized with the time $t$. On the  other hand the lattice models (LERW, HE and PF) usually have a natural parametrization given by the number of steps with equal length along the curve. In general the scaling limit of the lattice models are not the same as SLE$_\kappa$, where this difference comes from parametrization of each model \cite{KennedyNatural1}. We will use from an appropriate method to re-parametrize the SLE$_\kappa$ curve.

\subsection{Natural parametrized SLE }

First step in simulating SLE is time-step discretization. 
Let us consider a partition of the time interval $[0, t]$, where it is  discretized into $0 = t_0 < t_1 < t_2 < \dots < t_n=t$. One method to simulate SLE is the foregoing approximation with the equally spaced discrete time points $t_i=idt$. In this method the points $z_i$ on the curve $\gamma(t)$ is given by an iteration process 
$ z_i=f_1\circ f_2 \circ \dots \circ f_j (\xi_j)$, where $f_j(z)=\sqrt{(z-\xi_j)^2-4dt}+\xi_j$ is the inverse conformal map and $\xi_j$ is the discretized drift, where it will be approximated by a piece-wise constant function in the uniform
partition of the time interval $[(i-1)dt,idt]$. Notice that the conformal map $f_i (\xi_i)$ can produce a small slit at $\xi_j$ with length $L_i =$ Im$(f_i(\xi_i)) = 2\sqrt{t_{i+1}-t_{i}}$ on the upper-half plane. In this method the two-dimensional distances $l_i = \mid\gamma(t_i)-\gamma(t_{i-1})\mid$ are extremely non-uniform \cite{kennedy,KennedyNatural1,Gherardi}. 

We hereby, require the natural parametrized SLE$_\kappa$ (N-SLE$_\kappa$) curve, where it is the discrete SLE$_\kappa$ curve $\{\gamma_i\}$ with an approximately equal step length $\mid\gamma(t_i)-\gamma(t_{i-1})\mid\ \approx \lambda$. There are some mathematical and numerical procedures used to find a sensible definition of N-SLE$_\kappa$ \cite{kennedy,Gherardi,Lawler1} (see appendix C).

\subsection{Subordinated SLE }

In order to understand the scaling relations for the models of diffusion with waiting time on the self-similar traces and also two sided diffusion on the fractal curves, we present here the subordinated version of SLE$_\kappa$.  

The motivation of our approach comes from this idea that the probability distribution of the point at the tip of the SLE$_\kappa$ trace satisfies Fokker-Planck equation (FPE)  \cite{Gruzberg} which basically one can think about it as the FPE of the position of a particle in the fractal interface. In a similar way one may think about the FPE for the probability distribution  of the tip of the N-SLE and also subordinated N-SLE (NS-SLE) curve in the continuum limit and study diffusion. In principle it should be possible to calculate analytically MSD and MFPT for these semi-1D interfaces  by using the FPE in two dimensions. Unfortunately we do not know how to write the FPE of the N-SLE and NS-SLE, therfore we just calculate numerically the scaling properties of the tip of NS-SLE as the diffusion process on the fractal paths. We will show that the scaling behaviors of the subordinated N-SLE are similar to the lattice models. 

For normal SLE given by Eq. (\ref{Loewner_equation}) the time variable is deterministic but we would like to set this variable
as an internal parameter $\tau$ which is also stochastic and strictly non-decreasing, this is called subordinating the process by the inverse time $\alpha$-stable subordinator $S^\alpha_{t}$ (see appendix. B). 

Using the above definition one can consider Loewner's map with the new time as $g_{S^\alpha_{t}}(z)$ which is still scale invariant in the following sense: the conformal map $\tilde{g}_{S^\alpha_{t}}(z)=\frac{1}{\lambda^{\alpha/2}}g_{S^\alpha_{\lambda t}}(\lambda^{\alpha/2}z)$
with $\tilde{B}(S^\alpha_{t}):=\frac{1}{\lambda^{\alpha/2}}B(S^\alpha_{\lambda t})$ satisfies the same Loewner equation as $g_t(z)$. The above scale invariance enforces scale invariance of the curve. 

We have discussed the simulation of SLE and N-SLE in the last section. The simulation of subordinated SLE (S-SLE) and natural parametrized subordinated SLE (NS-SLE) are similar. The only difference is the conformal map  $f_j(z)=\sqrt{(z-\xi(S^{\alpha}_{t_j}))^2-4dS^{\alpha}_{t_j}}+\xi (S^\alpha_{t_j})$ where $dS^\alpha_{t_n}=S^\alpha_{t_n}-S^\alpha_{t_{n-1}}$ and $0 = t_0 < t_1 < t_2 < \dots < t_n=t$. The time steps $\Delta_j=dt$ ($t_i=\sum^i_{j=1}\Delta_j$) in the case of S-SLE$^{\alpha}_\kappa$ are selected uniformly and in the NS-SLE$^{\alpha}_\kappa$ the non-uniform time steps $\Delta_i$ are computed by using Jacobian scheme. The only difference between S-SLE$^{\alpha}_\kappa$ (NS-SLE$^{\alpha}_\kappa$) and normal SLE$_\kappa$ (N-SLE$_\kappa$) is in the growth process of each of them. In the first case the tip of the curve has waiting time according to the $\alpha$-stable Levy process.

As discussed earlier, the scaling exponents $\nu$ and $\beta$ for subordinated walk along discrete fractal interfaces are defined 
explicitly in Eq. (\ref{subordinatedscalingX2}) where they are in agreement with numerical simulations. We will consider a tip of NS-SLE$^{\alpha}_\kappa$ as a subordinated growth process where it is a mathematical model for subordinated walk along fractal curves. The scaling exponents of this subordinated process is collected in Table. ~\ref{tab3}, where they are in a good agreement with Eq. (\ref{subordinatedscalingX2}) and also with the numerical simulation of the subordinated random walk along the self-similar discrete curves (see Table. ~\ref{tab2}). 

 \begin{table}[htp]
\begin{center}
\begin{tabular}{ccccc}\hline\hline

NS-SLE$^{\alpha}_\kappa$ &  $\nu(\alpha=0.8)$  & $\beta(\alpha=0.8)$ & $\nu(\alpha=0.9)$  & $\beta(\alpha=0.9)$ \\
\hline
$\kappa=2.0$ &$1.28 \pm {0.02}$ & $1.52 \pm {0.05}$&$1.42 \pm {0.04}$ & $1.35 \pm {0.05}$\\
$\kappa=4.0$  &$1.02 \pm {0.05}$ & $1.84 \pm {0.04}$&$1.16 \pm {0.05}$ & $1.63 \pm {0.05}$\\
$\kappa=6.0$  &$0.92 \pm {0.03}$ &$2.15 \pm {0.05}$&$1.00 \pm {0.03}$ & $1.90 \pm {0.06}$   
\\ \hline\hline
\end{tabular}
\end{center}
\caption{\label{tab3} Numerical values of the scaling exponents $\nu$ and $\beta$ for the tip of the NS-SLE$^{\alpha}_\kappa$ curves with $\kappa=$2.0, 4.0 and 6.0 and $\alpha$=0.8 and 0.9.}
\end{table}  

We also notice that another way to subordinate the forcing function in Eq. (\ref{Loewner_equation}) is based on the iterated Brownian motion \cite{Iterative}. Consider two stochastic processes $B_t$ and $Y^{H}_t$, where the first one is the Brownian motion and the second one is the fractional Brownian process. The iterated Brownian process is defined as $B({|Y^{H}_t|})$, where ${|Y^{H}_t|}$ corresponds to the non-negative random time. It is easy to verify that the fractional Brownian time Brownian motion $B({|Y^{H}_t|})$ is a self-similar process of index $H/2$, that is, for any $\lambda$
\begin{equation} \label{brownian_time_scaling} 
B({|Y^{H}_{\lambda t}|})=B(\lambda^H{|Y^{H}_t|})=\lambda ^{H/2}B({|Y^{H}_t|}).
\end{equation}

Simulation of the the natural parametrized version of fractional Brownian time SLE (NF-SLE$^{H}_\kappa$) is similar to the NS-SLE case.  First consider discrete times $0 = t_0 < t_1 < t_2 < \dots < t_n=t$  and the conformal map $f_j(z)=\sqrt{(z-\xi(|Y^{H}_{t_j}|))^2-4d|Y^{H}_{t_j}|}+\xi (|Y^H_{t_j}|)$, where the infinitesimal values of the local time $d|Y^{H}_{t_j}|=|Y^{H}_{t_j}|-|Y^{H}_{t_{j-1}}|$ can get positive and negative values. The length of NF-SLE$^{H}_\kappa$ curves increases for $d|Y^{H}_{t_j}|>0$ and decreases for $d|Y^{H}_{t_j}|<0$. This dynamical process is very similar to the two sided diffusion on the lattice fractal models. Our estimations for the two scaling parameters $\nu$ and $\beta$ for MSD and MFPT, Table ~\ref{tab4}, are in a good agreement with the predicted values in Eq. (\ref{twosidedscalingX2}) and also numerical results coming from the lattice models (see Table ~\ref{tab1}).
\begin{table}[htp]
\begin{center}
\begin{tabular}{ccccc}\hline\hline

NF-SLE$^{H}_\kappa$ &  $\nu(H=0.8)$  & $\beta(H=0.8)$ & $\nu(H=0.9)$  & $\beta(H=0.9)$ \\
\hline
$\kappa=2.0$ &$1.28 \pm {0.01}$ & $1.57 \pm {0.01}$&$1.44 \pm {0.01}$ & $1.40 \pm {0.02}$\\
$\kappa=4.0$  &$1.05 \pm {0.03}$ & $1.90 \pm {0.02}$&$1.16 \pm {0.05}$ & $1.70 \pm {0.04}$\\
$\kappa=6.0$  &$0.90 \pm {0.03}$ &$2.20 \pm {0.02}$&$1.01 \pm {0.03}$ & $2.00 \pm {0.05}$   
\\ \hline\hline
\end{tabular}
\end{center}
\caption{\label{tab4} Numerical values of the scaling exponents $\nu$ and $\beta$ for the tip of the NF-SLE$^{H}_\kappa$ curves with $\kappa=$2.0, 4.0 and 6.0 and $H$=0.8 and 0.9.}
\end{table}

\section{Conclusion}

To conclude, we  studied the diffusive dynamics of the random processes on the self-similar curves and measured the scaling exponents of mean squared displacement and mean first passage time expressed in Eq. (\ref{GenscalingX2}). The various scaling exponents for MSD and MFPT
are obtained numerically for  two sided diffusion and diffusion with waiting time on three discrete fractal curves, i.e. loop-erased random walk, harmonic explorer and percolation front. It appears that the
exponents only depend on the fractal dimension $d_f$ of the curves and the scaling exponent $H$ for the two sided diffusion and $\alpha$ for the subordinated diffusion. 

Finally, we rederived the properties of the anomalous diffusion (FPT, MSD) on the discrete fractal curves  with subordinated version of the natural parametrized SLE. Our results offer a new method to investigate diffusion in the fractal interfaces. We believe that these results is a starting point for
the development of the subordinated version of SLE.

\section*{Appendix. A}
To find harmonic explorer and percolation front, we used overruled harmonic explorer process  on the very large rectangular domain.
This domain on the upper half plane as shown in Fig.~4 is splitting into three parts, a left boundary with yellow condition and a right boundary with blue and also uncoloured inner part. This boundary condition is used to limit the EP path to this part of half-plane to start from $r_0$ and stop when reach to $r_*$ or length $N$. The explorer process is the unique path from the origin. In each step, there is a yellow hexagon on the left and blue one on the right \cite{kager}.

\begin{figure}
\begin{center}
\includegraphics[width=.7\linewidth]{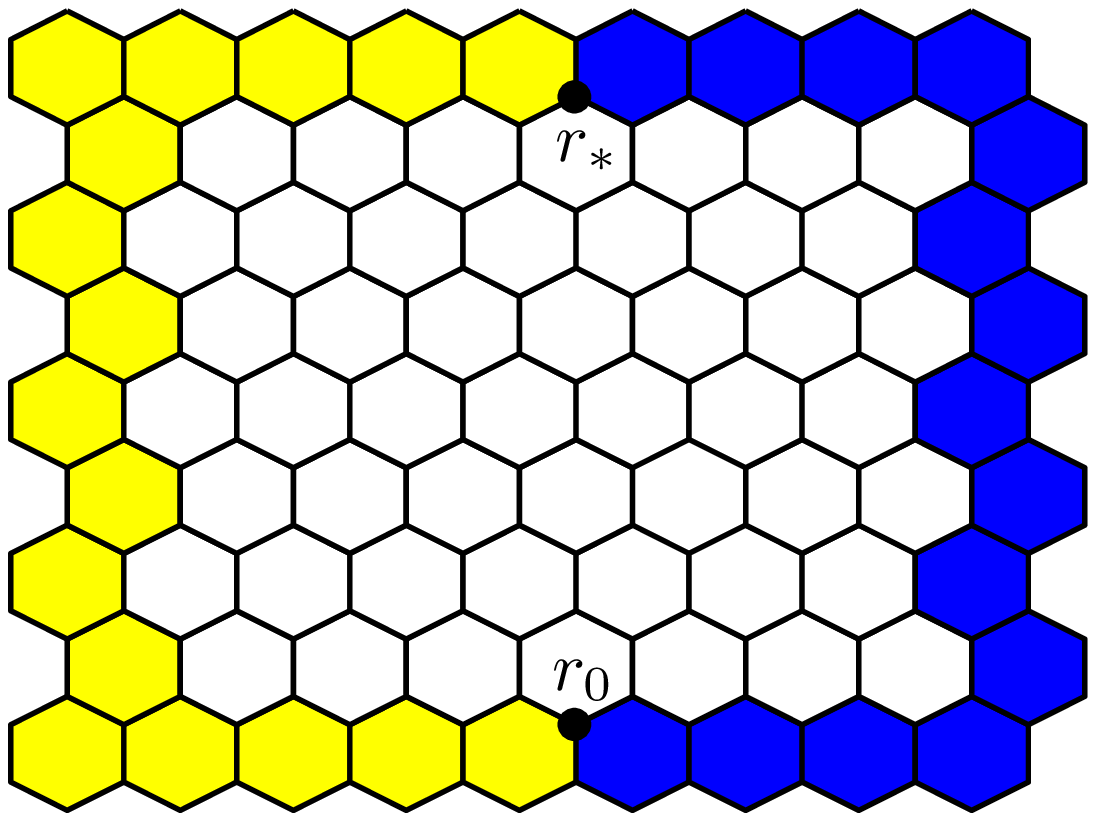}
\includegraphics[width=.7\linewidth]{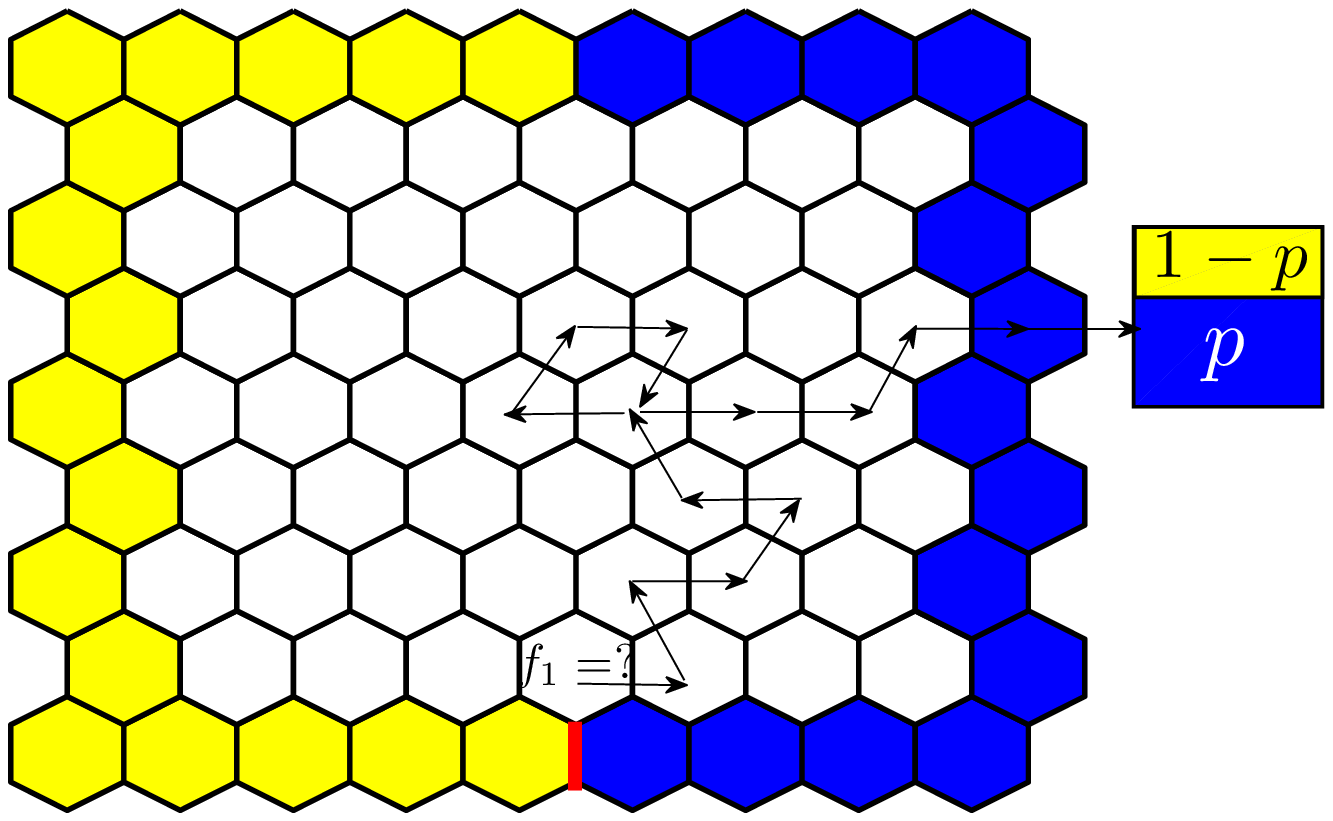}
\includegraphics[width=.7\linewidth]{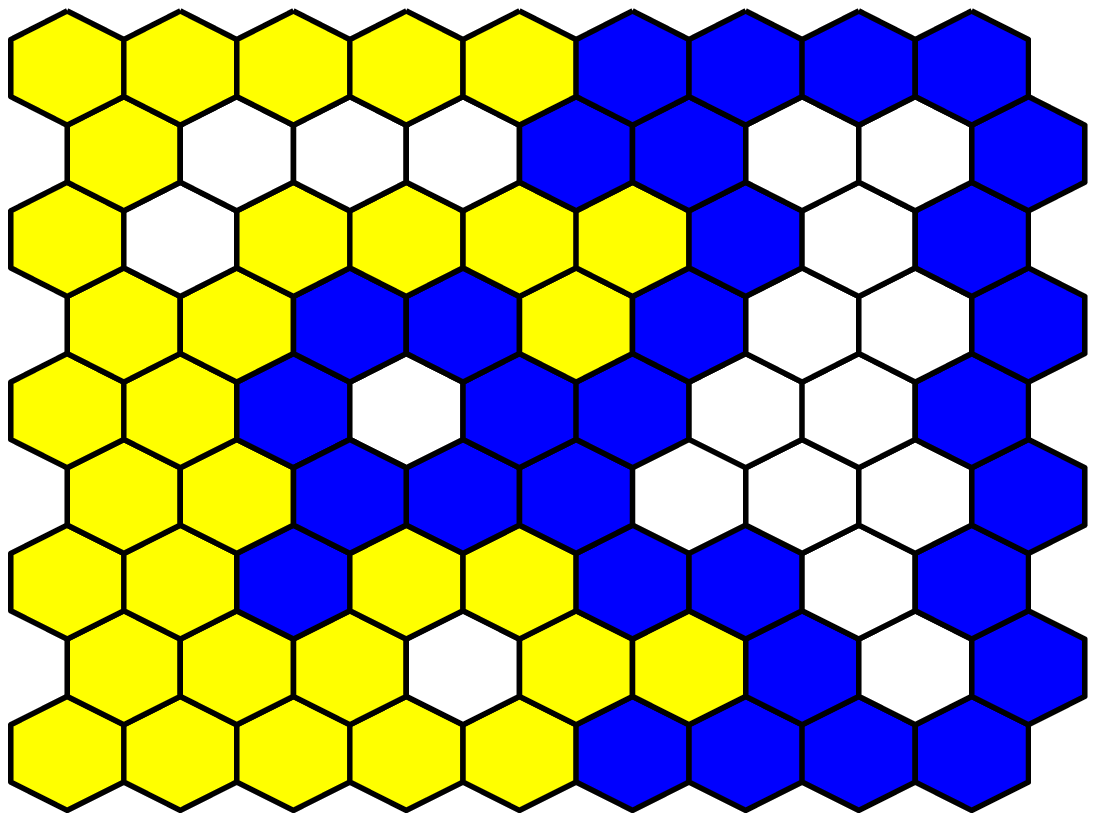}
\caption{\label{4} A rectangular domain with appropriate boundary conditions is used to build an explorer path. Top: Three parts of the domain which consist of left (right) boundary with yellow (blue) hexagons and uncoloured hexagons. Middle: First step to identify colour of face $f_1$. A random walker moves around uncoloured sites to hit the boundary. In this example, the colour of face $f_1$ with probability $p$ ($1-p$) will be blue (yellow). Notice that walker turn to right (left) when the yellow (blue) is selected. Bottom:  A complete exploration process in a rectangle. } \end{center}
\end{figure}

To generate this path dynamically, a growth process starts from the point $r_0$ on the lower boundary. In the first step the colour of face $f_1$ in front of $r_0$ is chosen so that to make it blue or yellow and the explorer is forced to turn left or right, respectively. To choose colour of face $f_1$, a random walker starts from $f_1$ and it stops when it crosses the rectangle's boundary for the first time. Now, the colour of $f_1$ with probability $0<p<1$ is yellow if the touch boundary is yellow. Note that two stochastic operations are used to colour one hexagon, a random walker to find boundary colour and a coin to does or does not accept the boundary colour. The new tip of explorer path is located in the position $r_1$ and a new face ($f_2$) should be coloured with the same restriction. In particular, the outcome of explorer process with $p=\frac{1}{2}, 1 $  as shown in Fig.~5 are percolation front and harmonic explorer, respectively. The fractal dimension of the overruled harmonic explorers has a linear relationship with $p$ and it is conjectured to be $d_f=2-\frac{p}{2}$ \cite{celani}. 

\begin{figure}
\begin{center}
\includegraphics[width=.8\linewidth]{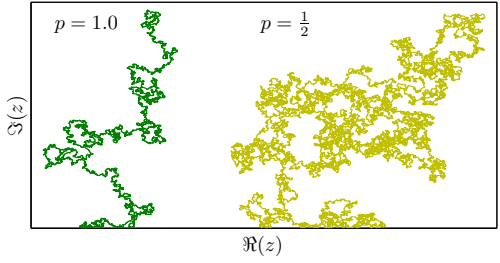}
\caption{\label{5} The \textit{overruled harmonic explorer} path with length $N=3\times 10^4$, Left: Harmonic explorer with $p=1.0$ and fractal dimension  $d_f=3/2$, Right: Percolation explorer with $p=1/2$ and fractal dimension $d_f=7/4$. } \end{center}
\end{figure}

\section*{Appendix. B}
The celebrated subordinated random time $S^{\alpha}_t$, is given by the Eq. (\ref{subordinator}) where it can be
efficiently generated by the algorithm proposed in \cite{Magdziarz}. As mentioned earlier the $\alpha$-stable subordinator $U(\tau)$ denotes the strictly increasing Levy motion with Laplace transform $\langle e^{-kU(\tau)}\rangle =e^{- \tau k^{\beta}}$ \cite{Janicki}. The first step in simulating $S^{\alpha}_t$ begins with approximating of the strictly increasing
 $\alpha$-stable Levy motion  $U(\tau)$ on the discrete times $\tau_i=i\Delta \tau$ ($i=0, 1, \dots, M$). The numerical integration of the process $U(\tau)$ for $0<\alpha\leq 1$ yields
 \begin{equation} \label{U_tau} 
U(\tau_{j+1})=U(\tau_j)+\Delta \tau ^{1/\alpha}L_{\alpha}(\beta),
\end{equation}
where $L_{\alpha}(\beta)$ is a Levy stable random variable with parameter $\beta$ and $U(0)=0$. We  use  skewed Levy-stable distribution ($\beta=1$), to ensure $U(\tau)$ gets almost increasing random process \cite{Janicki}. It can be generated by

\begin{equation} \label{Levy} 
L_{\alpha}(1.0)=\frac{\sin[\alpha(V+\frac{\pi}{2}) ]}{ [\cos(V)] ^{1/\alpha}}\times \{\frac{\cos[V-\alpha(V+\frac{\pi}{2})]}{W}\}^{(1-\alpha)/\alpha},
\end{equation}
where $V$ is a random variable with uniform distribution between $(-\frac{\pi}{2},\frac{\pi}{2})$ and $W$ has exponential distribution with mean 1. For the time horizon $T$, the summation process in Eq. (\ref{U_tau}) ends when we get $U(\tau_{M-1})\leq T<U(\tau_M)$. One can observe that $U(\tau)$ is strictly increasing and $M$ always exists \cite{Magdziarz}.

Now, for every $t_i\in (0=t_0<t_1<t_2\dots<t_N=T)$, we find $\tau_j$ such that $U(\tau_{j-1})< t_i\leq U(\tau_j)$, and from the definition in the Eq. (\ref{subordinator}), we can define $S^{\alpha}_{t_i}=\tau_j$. From Eqs. (\ref{U_tau}) and (\ref{Levy}) it is clear that $L_{\alpha}(1)=1$ and $S^{\alpha}_{t_i}=t_i$ in the $\alpha \rightarrow 1$ limit, where at this limit subordinated time converges to the normal time. 

\section*{Appendix. C}

A standard procedure \cite{Gherardi} to find the half plane N-SLE$_\kappa$ trace is based on a change in the size of the $i^{th}$ slit length $L_i$ (It is a function of time step parameter $\Delta_i=t_{i}-t_{i-1}$ as $L_i=2\sqrt{\Delta_i}$), by the Jacobian. The Jacobian $\mid J_{i-1}\mid\approx  \mid(\xi_i-\xi_{i-1})G^{''}_{i-1}(\xi_{i-1})\mid $
of the conformal map $G_i=f_1\circ f_2 \circ \dots \circ f_i $ acts on the corresponding segment to rescale the  length $L_i$ for the $i^{th}$ slit by
\begin{equation} \label{rescale1} 
L_i\approx \frac{\lambda}{\mid J_{n-1}\mid},
\end{equation}
where $\lambda>0$ is the step length. For a piece-wise constant Brownian process  $\xi_i=\xi_{i-1}\pm \sqrt{\kappa \Delta_i}$ (the sign of $\sqrt{\kappa \Delta_i}$ is chosen randomly according to the uniform probability distribution), the above approximation yields
\begin{equation} \label{rescale2} 
\Delta_i=\frac{\lambda}{2\sqrt{\kappa}\mid G^{''}_{i-1}(\xi_{i-1})\mid}.
\end{equation}
Note that computing points along  the N-SLE$_\kappa$ curve requires this adaptive choice of  $\Delta_i$  and the total time with these non-uniform time steps will be equal to $t_i=\sum^{i}_{n=1} \Delta_i$. In this procedure the distances between two sequential points $l_i\approx \lambda$ approximately remain constant. 

In our study we followed one straightforward motivation for computing the Jacobian. If one consider $h_i(z)=f_i(z+\xi_i)-\xi_{i-1}$.
The conformal map $h_i(0)$, maps the upper half plane onto the upper half plane plus a slit. The length of this slit equals to $2\sqrt{\Delta_i}$ and the position on the real line equals to $\delta_i=\xi_i-\xi_{i-1}$ . Following a simpler strategy one can decompose the incremental map $h_i(z)$ to $h_i(z)=T_{\delta_i} \circ \phi_i^{\mathbb{H}}$, where $\phi_i^{\mathbb{H}}=\sqrt{z^2-4\Delta_i}$ is the slit map and $T_{\delta_i}(z)=z+\delta_i$ is a translation map by the real value $\delta_i$. The $i^{th}$ points of the the SLE or N-SLE curve computed from 
$ \gamma(t_i)=g_n(0)$ where
\begin{equation} \label{confmapg} 
 g_n(z)=T_{\delta_1} \circ \phi_1^{\mathbb{H}} \circ  T_{\delta_2} \circ \phi_2^{\mathbb{H}} \circ \dots \circ T_{\delta_i} \circ \phi_i^{\mathbb{H}}(z).
\end{equation}
 We now consider a new format of Eq. (\ref{rescale2}) as
\begin{equation} \label{rescale3} 
\Delta_i=\frac{\lambda}{2\sqrt{\kappa}\mid g^{''}_{i-1}(0)\mid}.
\end{equation}
 where
 \begin{equation} \label{gzeg} 
 g^{''}_{i}(0)=\mid \phi^{''}_n(0) \mid \prod^{n-2}_{j=0} \mid \phi^{'}_{n-1-j}(\Gamma_j) \mid.
\end{equation}
In the above equation a $\Gamma_j$ is defined as
\begin{equation} \label{gammaj} 
\Gamma_j=T_{\delta_{n-j}} \circ \phi_{n-j}^{\mathbb{H}} \circ  T_{\delta_{n-j+1}} \circ \phi_{n-j+1}^{\mathbb{H}} \circ \dots \circ T_{\delta_n} \circ \phi_n^{\mathbb{H}}(0).
\end{equation} 
Following \cite{Gherardi} the proposed method of approximating sample paths of N-SLE$_\kappa$ $\{\gamma(t_i), i=0,\dots,N \}$, consists of six steps: ($\textbf{1}$) Set the constants $\lambda$, $\kappa$ and $N$. (\textbf{2}) Set $n=1$ and $\Delta_1=1$. (\textbf{3}) Compute $\sqrt{\kappa\Delta_n}$ according to steps \textbf{1} and \textbf{2} with a random sign ($\pm$) with equal probability. 
 (\textbf{4}) Calculate $\gamma(t_n)=g_n(0)$ using the iteration map as we said in  Eq. (\ref{confmapg}). 
 (\textbf{5}) Compute $\Delta_{n+1}$ using Eq. (\ref{rescale2}), (\ref{gzeg}) and (\ref{gammaj}). 
 (\textbf{6}) If $n<N$ increase $n$ by one and repeat steps \textbf{3} to \textbf{6}. The
typical curves of the N-SLE$_\kappa$ for $\kappa=$2, 4 and 6 are presented in Fig. ~6.

\begin{figure}
\begin{center}
\includegraphics[width=0.9\linewidth]{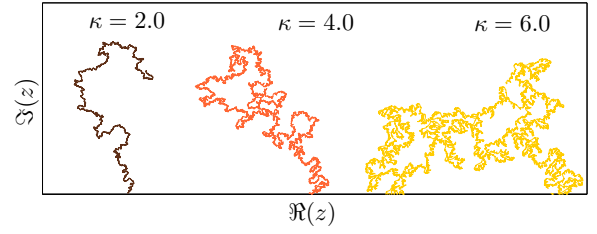}
\caption{\label{6} The N-SLE$_\kappa$ curves with length $N=1\times 10^4$, $\lambda=0.001$ and $\kappa$=2.0, 4.0 and 6.0 from left to right. } \end{center}
\end{figure}

\section*{Acknowledgements}

M. G. Nezhadhaghighi kindly acknowledge intensive discussions with Marco Gherardi.

\end{document}